# Beam Orientation Optimization for Intensity Modulated Radiation Therapy using Adaptive $l_1$ Minimization

**Xun Jia[1], Chunhua Men[1], Yifei Lou[2], and Steve B. Jiang[1]**

[1]Center for Advanced Radiotherapy Technologies and Department of Radiation Oncology, University of California San Diego, La Jolla, CA 92037-0843, USA

[2]Department of Mathematics, University of California Los Angeles, Los Angeles, CA 90095-1555, USA

E-mail: sbjiang@ucsd.edu

Beam orientation optimization (BOO) is a key component in the process of IMRT treatment planning. It determines to what degree one can achieve a good treatment plan quality in the subsequent plan optimization process. In this paper, we have developed a BOO algorithm via adaptive $l_1$ minimization. Specifically, we introduce a sparsity energy function term into our model which contains weighting factors for each beam angle adaptively adjusted during the optimization process. Such an energy term favors small number of beam angles. By optimizing a total energy function containing a dosimetric term and the sparsity term, we are able to identify the unimportant beam angles and gradually remove them without largely sacrificing the dosimetric objective. In one typical prostate case, the convergence property of our algorithm, as well as the how the beam angles are selected during the optimization process, is demonstrated. Fluence map optimization (FMO) is then performed based on the optimized beam angles. The resulted plan quality is presented and found to be better than that obtained from unoptimized (equiangular) beam orientations. We have further systematically validated our algorithm in the contexts of 5-9 coplanar beams for 5 prostate cases and 1 head and neck case. For each case, the final FMO objective function value is used to compare the optimized beam orientations and the equiangular ones. It is found that, our BOO algorithm can lead to beam configurations which attain lower FMO objective function values than corresponding equiangular cases, indicating the effectiveness of our BOO algorithm.

## 1. Introduction

Intensity modulated radiation therapy (IMRT) is currently considered to be one of the most effective radiation therapy techniques for many clinical scenarios. The treatment planning process of IMRT is usually conducted by first selecting a given number of beam orientations and then optimizing the fluence maps at those beam angles. In this process, the first subproblem, known as beam orientation optimization (BOO), plays a vital role in determining the final treatment plan quality (Soderstrom and Brahme, 1995; Pugachev *et al.*, 2001; Bortfeld, 2010), especially when using a small number of beams or dealing with complicated anatomy. In those cases where a relatively large number of beams are used, the gain of BOO in terms of treatment plan quality becomes diminishing (Stein *et al.*, 1997). Yet, using many beams in IMRT could lead to problems such as prolonged treatment time and increased potential errors due to patient motion. Therefore, it is highly desirable to perform BOO to obtain plans with a small number of beams while achieving quality same as or even better than those plans with more but unoptimized beam angles.

In the current IMRT practice, the beam orientation selection is usually achieved via a cumbersome trial-and-error approach conducted by experienced treatment planners. Sometimes equiangular beams are also used in many clinical scenarios for simplicity. These approaches, however, usually result in suboptimal treatment plan qualities despite the carefully tuned algorithm in the subsequent stage of the fluence map optimization (FMO).

The BOO problem belongs to the category of combinatorial optimization, in that it searches for an optimal combination of a given number of beam orientations among all candidate angles to achieve the best plan quality. By enumerating all the possible beam orientation combinations, one can ensure the optimality of the selected beam orientations. Nonetheless, the associated computational time of this exhaustive searching strategy is extremely long, making it practically infeasible. In the past, many research groups have been actively working on designing efficient algorithms to automate the BOO process in IMRT. For example, heuristic approaches (Bortfeld and Schlegel, 1993; Lu *et al.*, 1997; Haas *et al.*, 1998; Rowbottom *et al.*, 2001; Pugachev and Xing, 2002; Hou *et al.*, 2003; Djajaputra *et al.*, 2003; Das *et al.*, 2003; Li *et al.*, 2004; Schreibmann *et al.*, 2004; Li and Yao, 2006), such as genetic algorithms or simulating annealing, have been developed to search for the optimal solution. However, besides the prolonged computation time due to the large searching space, it is very hard for these heuristic searching strategies to assert the global optimality of the solution due to the existence of local minima.

On the other hand, a number of other work utilize predetermined criteria for geometrical or dosimetric considerations to rank the candidate beam orientations and those with highest ranking scores are selected (Woudstra and Storchi, 2000; Pugachev and Lei, 2001; Meedt *et al.*, 2003; Woudstra *et al.*, 2005; Meyer *et al.*, 2005; Vaitheeswaran *et al.*, 2010). Though the algorithms of such kind can be quite computationally efficient, this sequential way of selecting beam orientations cannot guarantee the optimality of the resulted beams due to the interplay between them.

In this paper, we will present our recent development towards the BOO problem by employing the idea of sparse optimization. Generally speaking, the sparse optimization problem attempts to retrieve a signal which is known *a priori* to be sparse, *i.e.* it contains only a small number of components of all the available candidates. We can immediately notice the similarity to our BOO problem, where we try to select only a small number of beam orientations among all the candidate angles to yield a high quality treatment plan. In solving problems of this kind, the concept of sparsity is of crucial importance. The main difficulties are, first, how to encode the sparsity into the optimization model in a mathematical language and second, how to tackle the designed optimization model. Recently, with the vast and fruitful development in the field of sparse optimization, these difficulties have been successfully overcome in a number of different mathematical contexts. It is therefore ready to bring the sparsity idea into the BOO context to provide a brand new and effective perspective for solving this problem.

As such, in the rest of this paper, we will develop and validate a BOO algorithm via the so called $l_1$-minimization approach, the most widely used approach for solving a sparse recovery problem. Specifically, we will introduce a sparsity energy term into our model which contains weighting factors for each beam angle adaptively adjusted during the optimization process. Such an energy term favors those solutions with only a few number of beam angles. By minimizing an energy function with this sparsity term and a dosimetric term, we can enforce that only a small fraction of total candidate beam angles are selected to achieve the dosimetric goal. The rest of this paper is structured as following. In section 2, we will present our BOO model as well as the optimization algorithm. Section 3 is devoted to the comprehensive validation of the model in a number of clinical cases. Finally, we conclude our paper in section 4 and provide some further discussions.

## 2. Methods

### 2.1 Background

Let us consider $l_p$-norm of a vector $x$ defined as $\|x\|_p = \left[ \sum_i |x_i|^p \right]^{1/p}$ with $p > 0$. One example is the $l_2$-norm, which is also known as Euclidean norm. One can also define the so called $l_0$-norm by taking a limit $p \rightarrow 0$, which counts the number of non-zero elements of $x$. Suppose we are looking for a solution for an over-determined linear system $Ax = b$ via least square method. If we have the knowledge *a priori* that we prefer a sparse solution $x$, namely it contains very few number of non-zero elements, it is straightforward to impose this condition in the solution process by seeking for a solution to an optimization problem (P1) $\min_x \|Ax - b\|_2^2 + \mu \|x\|_0$, where $\mu > 0$ is a parameter representing the relative weights between the two terms. This minimization problem explicitly gives penalization according to the number of non-zero elements of $x$, hence yielding a sparse solution. The parameter $\mu$ controls to what degree we would like to enforce the sparsity. However, the energy function considered here is not convex due to the $l_0$-norm term. As a consequence, exactly solving this problem again requires an exhaustive searching strategy akin to the BOO problem. An alternative way of approaching this problem is to replace the non-convex $l_0$-norm by an $l_1$-norm, which is being the convex norm closest to $l_0$-norm. This substitution leads to solving the problem (P2) $\min_x \|Ax - b\|_2^2 + \mu \|x\|_1$. In many contexts, it has been shown that this utilization of $l_1$-norm is very effective in terms of enforcing sparsity on the solution. Moreover, computation-wise, solving an $l_1$-minimization problem is much easier due to the convex form of the energy function and many advanced algorithms invented.

Note that the problems (P1) and (P2) are not completely equivalent mathematically. Generally, it is the problem (P1) that one would like to solve, as it directly penalizes the number of non-zero elements in the solution. Yet, the use of the $l_1$-norm is found to be effective to produce a sparse solution, but not necessarily the sparse solution as in problem (P1). The substitution of the $l_0$-norm by an $l_1$-norm is merely a practical way of making the sparsity problem tractable.

### 2.2 Beam Orientation Optimization Model

We will denote beamlet intensity by $x_j^\theta$, where $j$ is an index of the beamlet within a beam angle $\theta$ . Given the so-called dose deposition coefficients $D_{ij}^\theta$, namely the dose received by voxel $i$ from the beamlet $j$ in the angle $\theta$ at its unit intensity, we can compute the voxel dose $z_i$ at a voxel $i$ using a linear model

$$z_i = \sum_{j,\theta} D_{ij}^\theta x_j^\theta, \tag{1}$$

Equivalently, we can rewrite Eq. (1) in a compact form $z = Dx$, where $z$ and $x$ are vectors consisting of $z_i$ and $x_j^\theta$, respectively, and $D$ is the dose deposition matrix with entries $D_{ij}^\theta$.

In our model we employ treatment plan evaluation criteria that are quadratic one-sided voxel-based penalties, *i.e.* the energy function at a voxel $i$ is chosen as

$$E_i[x] = \alpha_i (\max\{0, T_i - z_i(x)\})^2 + \beta_i (\max\{0, z_i(x) - T_i\})^2, \tag{2}$$

where $\alpha_i$ and $\beta_i$ represent the penalty weights for underdosing and overdosing penalty, respectively. Specifically, for critical structures, $\alpha_i = 0$ and $\beta_i > 0$, so that there is penalty only when the received dose exceeds the threshold $T_i$. In contrast, for target structures, $\alpha_i > 0$ and $\beta_i > 0$, so that there is a penalty as long as there is deviation between the received dose and the prescription $T_i$. Generally speaking, the parameters $\alpha_i$ and $\beta_i$ can vary from voxel to voxel. In this work, they are chosen as organ dependent for simplicity, *i.e.* those voxels in a same organ will have same parameters. Moreover, the values of these parameters are set empirically as in our previous work (Men *et al.*, 2009). $E_i[x]$ is a function of the beamlet intensities $x$ due to the relationship between $x$ and $z$. The total dosimetric energy function is simply a summation of $E_i[x]$ over all voxels, namely, $E_{Dose}[x] = \sum_j E_i[x]$ . Minimizing this term will enforce a desired dose distribution.

As for the beam orientation optimization, it is our goal to select only a few beam angles which can effectively minimize the energy $E_{dose}[x]$. Or equivalently, we are searching for a solution which has a sparse representation at the level of beam orientation. In an optimization approach, this can be achieved by minimizing an energy function of a form of $l_1$-norm only at the angular level, but not at the beamlet level within each beam angle. As such, we consider an energy function of a $l_{2,1}$-norm defined as

$$E_{Angle}[x] = \sum_\theta \mu_\theta [\sum_j x_j^{\theta^2}]^{1/2} = \sum_\theta \mu_\theta \|x^\theta\|_2, \tag{3}$$

where $x^\theta = \left(x_1^\theta, x_2^\theta, \dots\right)^T$ is the vector consisting of beamlet entries only within a given angle $\theta$. The energy defined in such a way groups all the beamlet intensities at the angle $\theta$ utilizing an $l_2$-norm within the angle and then takes $l_1$-norm at the beam angle level. By minimizing such an energy term, we can ensure the sparsity at only the beam angle level but not at the beamlet level within each angle.

In considering the dosimetric objective and the sparsity objective, we propose to achieve the beam orientation optimization by minimizing a total energy $E[x] = \mu E_{Dose}[x] + E_{Angle}[x]$ with respect to the beamlet intensity vector $x$, subject to the condition $x_j^\theta \geq 0$. A constant $\mu > 0$ is chosen to adjust the relative weights between the dosimetric term and the angular sparsity term. Note that this $\mu$ is different from those $\mu_\theta$ factors in Eq. (3), which specify the relative weights between beam angles. The discussion regarding to the choice of $\mu_\theta$ will be presented in Section 2.4.

It is worth mentioning that the underlying assumption of using Eq. (3) to enforce sparsity is that the importance of an angle $\theta$ can be characterized by the associated $\|x^\theta\|_2$. Note that it is the product of the dose deposition matrix $D^\theta$ and the fluence map vector $x^\theta$, namely $D^\theta x^\theta$, that describes the contribution of the angle $\theta$ to the desired dosimetric objective. We have to normalize the dose deposition matrix $D^\theta$ to make it fair to use $\|x^\theta\|_2$ for comparing different beam angles. Therefore, in the following computation we have normalized the dose deposition matrix $D^\theta$, such that its Frobenius norm is unity, *i.e.* $\left[\sum_{i,j} D_{i,j}^{\theta^2}\right]^{1/2} = 1$.

### 2.3 Optimization Algorithm

Let us first consider the algorithm that optimizes the total energy $E[x]$ for a given set of $\mu_\theta$ factors. The total energy $E[x]$ is convex and therefore it is sufficient to consider its optimality condition:

$$0 = \frac{\partial E}{\partial x} = \mu \frac{\partial E_{Dose}}{\partial x} + \frac{\partial E_{Angle}}{\partial x}. \tag{4}$$

Let us split this condition by introducing an auxiliary vector $g$ and a scalar $\lambda$ as

$$0 = x - g - \lambda\mu \frac{\partial E_{Dose}}{\partial x}, \tag{5a}$$

$$0 = x - g + \lambda \frac{\partial E_{Angle}}{\partial x}, \tag{5b}$$

Noting that Eq. (5b) is the optimality condition of an energy $E_{Aux}[x] = \frac{1}{2}\|x - g\|_2^2 + \lambda E_{Angle}[x]$. The optimization can be performed by iterating two steps: 1) update $g = x - \lambda\mu\frac{\partial E_{Dose}}{\partial x}$ as indicated by Eq. (5a), and 2) solve a subproblem $\min_x E_{Aux}[x]$ as indicated by (5b). This algorithm is called forward-backward splitting and its mathematical correctness has been studied previously (Combettes and Wajs, 2005; Hale *et al.*, 2008). It is interesting to observe that the subproblem in step 2) has a close form solution. In fact, the vector that minimizes $E_{Aux}[x]$ can be expressed as

$$x^\theta = g^\theta \max\left(1 - \frac{\lambda\mu_\theta}{\|g^\theta\|_2}, 0\right), \tag{6}$$

where $g^\theta$ is defined by single out those elements of the vector $g$ corresponding to the angle $\theta$. The derivation of Eq. (6) is briefly shown in Appendix. The meaning of Eq. (6) is quite straightforward. After step 1), we obtain a vector $g^\theta$ via a gradient descent update for the energy $E_{Dose}$. This update, however, does not consider the sparsity energy term. To include this sparsity consideration into the optimization, Eq. (6) compares $\|g^\theta\|_2$ with a threshold $\lambda\mu_\theta$. In particular, if $\|g^\theta\|_2 < \lambda\mu_\theta$, the beam intensity profile at this angle is considered to be unimportant and is discard. By iteratively performing these two steps, we can keep selecting those angles of less importance and throw them away, leading to a solution $x^\theta$ which is sparse at the angular level.

In addition, since $x_j^\theta$ represent the beamlet intensity, which cannot be negative, we also impose this condition by simply truncating all the negative entries in the vector $x$ in each iteration step after step 1). In summary, the algorithm we employed to minimize the energy $E[x]$ is as following:

---
**Algorithm  A1:**

---
Initialize the vector $x$. Do the following steps till convergence

1.  Update: $g = x - \lambda\mu\frac{\partial E_{Dose}}{\partial x}$;

2.  Truncate: $x_j^\theta = 0$ if $x_j^\theta < 0$;

3.  Compute: $x^\theta = g^\theta \max\left(1 - \frac{\lambda\mu_\theta}{\|x^\theta\|_2}, 0\right)$.

---

Note that step 1 in A1 is merely a gradient descent step with respect to the energy $E_{Dose}[x]$ with a step size $\lambda\mu$. In practice, we have also performed an inexact line search (Bazaraa *et al.*, 2006) to adaptively select the parameter $\lambda$ in each iteration step in order to speed up the convergence. The convergence of this algorithm can be identified by comparing the energy function $E[x]$ at two successive iteration steps. Once the relative difference of $E[x]$ between two steps are smaller than a preset criteria, for instance 0.1%, the convergence is considered to be achieved.

### 2.4 Adaptively reweighting

The choice of the weights $\mu_\theta$ is apparently of crucial importance. In this work, we propose to generate the weighting factors $\mu_\theta$ adaptively according to the obtained $\|x^\theta\|_2$ during the optimization. In particular, we prefer to apply a large $\mu_\theta$ to those angles which are considered to be less important, or unlikely to be the optimal beam angles, so as to suppress their fluence maps. As for judging whether a given beam angle is important or not relative to others, there are two considerations. First, in terms of contributing to the dosimetric objective, the relative importance of beam angles between each other can be readily characterized by the quantity $\left\| x^\theta \right\|_2$. Therefore, we prefer to apply a large $\mu_\theta$ to those angles with small $\left\| x^\theta \right\|_2$. Second, the importance of a beam angle $\theta$ should be justified by comparing its $\left\| x^\theta \right\|_2$ value with only those at nearby angles, but not with all angles. This is because the $\left\| x^\theta \right\|_2$ value at an important beam angle may be larger than those at other angles close by, but may not be necessarily larger than those at some unimportant angels far away.

For the above two reasons, once given a set of $\left\| x^\theta \right\|_2$ at all angles $\theta$, we propose to generate the weighting factors $\mu_\theta$ as following. For a given angle $\theta$, 1) locate its two nearby beam angles $\theta_+$ and $\theta_-$ which have nonvanishing $\left\| x^\theta \right\|_2$, where cyclic boundary condition along beam angle direction is considered; 2) compute the maximum of $\left\| x^\theta \right\|_2$ among those three angles, and denote $\left\| x^\theta \right\|_2^{max} = \max \left( \left\| x^\theta \right\|_2, \left\| x^{\theta_+} \right\|_2, \left\| x^{\theta_-} \right\|_2 \right)$; 3) compute the weighting factor for the beam angle $\theta$ as

$$\mu^\theta = \exp \left[ -\left( \frac{\left\| x^\theta \right\|_2}{\left\| x^\theta \right\|_2^{max}} - 1 \right) \right]. \tag{7}$$

The weighting factor $\mu^\theta$ generated in such a way is unity for an angle which attains a locally maximum $\left\| x^\theta \right\|_2$ value, while yield a factor larger than unity otherwise. To illustrate this idea, a typical set of weighting factors $\mu^\theta$ are plotted in Fig. 1 along with the profile $\left\| x^\theta \right\|_2$ based on which $\mu^\theta$ are determined. In such an example, all beam angles with vanishing fluence map are assigned large weights ($\mu^\theta = 2.718$). In addition, the beam angle at $50°$ also has a large $\mu^\theta$ due to its small $\left\| x^\theta \right\|_2$ comparing to those at $25°$ and $60°$.

The adjustment of $\mu^\theta$ should not interfere with the optimization process described in Algorithm A1, since A1 is only correct with constant $\mu^\theta$ values. In practice, we alternatively perform the Algorithm A1 and the adaptive adjustment of $\mu^\theta$. Once a new set of $\mu^\theta$ is generated, the Algorithm A1 is conducted till its convergence before the next adjustment of $\mu^\theta$ is applied.

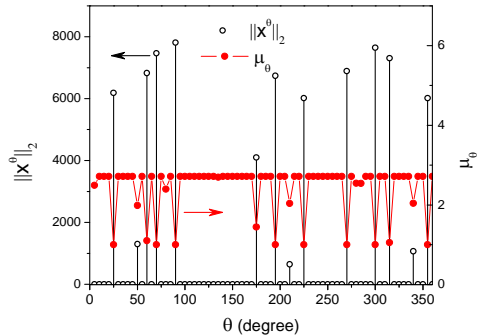

**Figure 1.** A typical form of $\left\| x^\theta \right\|_2$ and its corresponding $\mu^\theta$ as a function of the beam angle $\theta$.

During this process, the Algorithm A1 plays an role of sparsifying the beam angle profile $\left\| x^\theta \right\|_2$ under the guidance of the weighting factors $\mu^\theta$. It is able to discard those angles which are less important than others for dosimetric considerations. After that, the $\mu^\theta$ will be adjusted according to the importance of each angle. The updated $\mu^\theta$ will be used as a guidance for the Algoritthm A1 in the next iteration. By iterating these two steps, the number of beam angles which have nonvanishing fluence map keep decreasing. Therefore, we can start our BOO algorithm with all the candidate beam angles and perform the computation as described above. Once a preset target number of beams, $N_A$, is achieved, the overall algorithm terminates. This algorithm is summarized as following:

---

**Algorithm A2:**

Initialize $x_\theta$, set $\mu^\theta = 1$ for all $\theta$. Give a target number of beams $N_A$.

Repeat the following:

1. Sparsify the fluence map profile $\left\| x^\theta \right\|_2$ by using Algorithm A1;

2. Adjust weighting factor $\mu^\theta$ using Eq. (7);

3. Check the number of beams with nonzero fluence map. If it is larger than $N_A$, go back to step 1;

4. Output those beam angles with nonzero fluence map as the solution.

---

## 3. Experimental Results

### 3.1 A typical Prostate Case

We first test our algorithms on a typical prostate cancer case. The prescription dose to planning target volume (PTV) is 73.8 Gy. We used a beamlet size of 5×5 mm$^2$ and voxel size of 2.5×2.5×2.5 mm$^3$ for target and organs at risk (OARs). For unspecified tissue (*i.e.*, tissues outside the target and OARs), we increased the voxel size in each dimension by a factor of 2 to reduce the optimization problem size. The full resolution was used when evaluating the treatment quality (does volume histograms (DVHs), dose color wash, isodose curves, *etc.*). Dose deposition coefficients are generated for 6 MV coplanar beams at 72 candidate orientations equally spaced in a full rotation.

### 3.1.1 Convergence Properties

The first result we present is the convergence properties of our BOO algorithm. In this example, it is our objective to select $N_A = 7$ beam angles. In Fig. 1(a), we first plot the dosimetric energy $E_{Dose}$, the sparsity energy $E_{Angle}$, and the total energy $E$ as a function of the iteration steps. At the beginning stage, because of the existence of large number of candidate beam angles, the optimization algorithm can utilize all those beam angles to minimize $E_{Dose}$ and hence has plenty room to improve the dosimetric goal. This leads to a sharp decreasing trend in $E_{Dose}$. Later on, the effects of the sparsity constraint start to take place in the optimization and the algorithm gradually removes some unimportant beam angles. During this process, the dosimetric goal has to be sacrificed due to the gradually reduced number of beams, causing an increasing trend in $E_{Dose}$. Yet, our algorithm is so effective that it only discard those unimportant beam angles. This results in the degree of sacrificing dosimetric goal only to a minimal extent, leading to a very slow rise of $E_{Dose}$.

The total energy gradually increases during the iteration except at the initial stage, as opposed to a monotonic decrease behavior in a usual optimization problem. This is due to the adaptive adjustment of the weighting factors $\mu_\theta$. To see this in more detail, we plot a zoom-in view of the total energy $E$ in the insert of panel (a). There are some sudden jumps of its value during the iteration corresponding to the adjustment of the weighting factor. Between successive jumps, the value of $E$ decreases due to the optimization procedure in A1. Those sudden jumps can also be observed in the sparsity energy $E_{Angle}$ for the same reason.

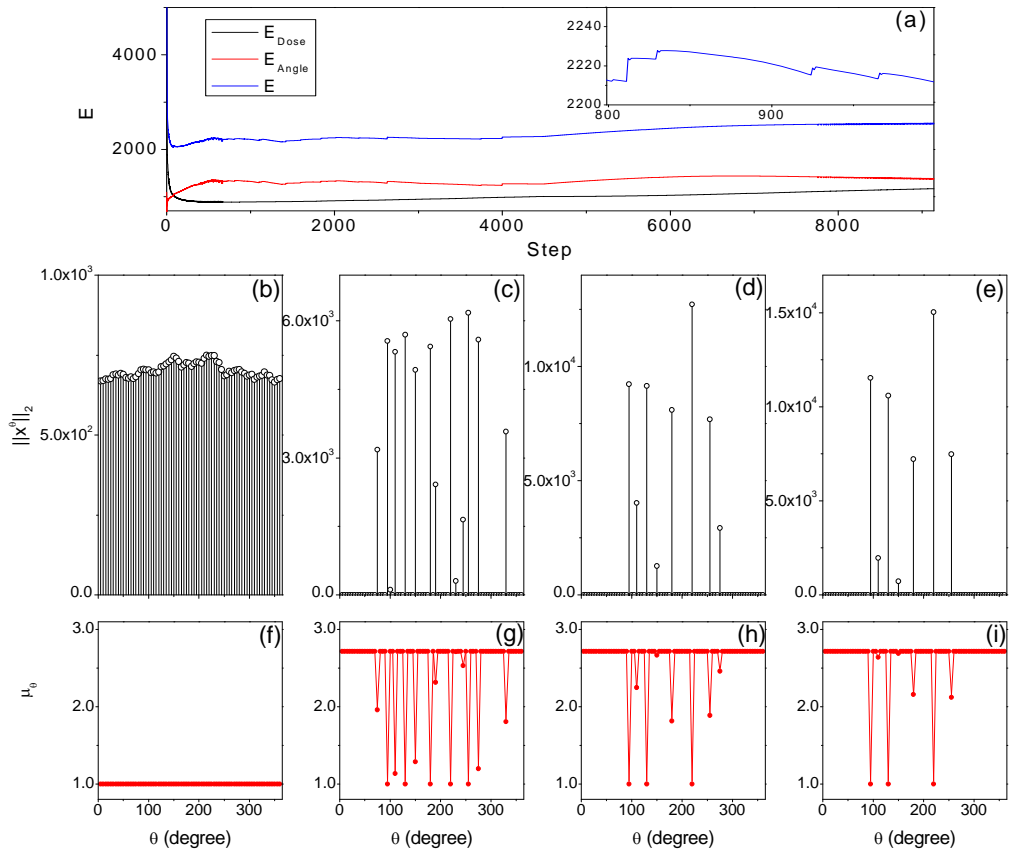

**Figure 2.** (a) The evolution of the energy functions during the iteration. The insert shows a zoom-in view of the total energy. (b)~(e) the quantity $\left\| x^\theta \right\|_2$ as a function of beam angles plotted at iteration steps 10, 3000, 7000, and 9514. (f)~(i) the quantity $\mu_\theta$ as a function of beam angles plotted at same iteration steps.

As for how the beam angles are selected during this iteration, we plot the profile $\left\| x^\theta \right\|_2$ as well as the weighting factors $\mu_\theta$ at iteration steps 10, 3000, 7000, and 9514. At beginning, such as in Fig 2 (b) and (f), all the beam angles participate into the optimization and the weighting factor $\mu_\theta$ at all angles are uniformly unity. As the

5    iteration continues, the sparsity and the adaptive adjustment of the weighting factors progressively show their effects. These two key components of our BOO algorithm can automatically identify those unimportant beam angles according to the $\left\| x^\theta \right\|_2$ profile and the adaptively adjusted weighting factors $\mu_\theta$. The fluence maps at those identified unimportant beam angles are then discarded in the algorithm A1. Finally, at step 9514 as

10   in Fig 2 (e) and (i), the number of beam angles with nonzero fluence maps meets our target number of beams $N_A = 7$, hence the BOO algorithm achieves its goal and terminates.

### 3.1.2 Compare with equiangular plans

To demonstrate the effectiveness of our algorithm, we compare the treatment plans obtained with and without the BOO in the case of $N_A = 7$ beams. For the plan with BOO, the orientations of these beams are generated by our algorithm. As indicated in Fig. 2 (e), these 7 angels are at $90\degree$, $105\degree$, $125\degree$, $145\degree$, $175\degree$, $215\degree$, and $250\degree$. As for the plan

20   without BOO, the angles are chosen to form an equiangular configuration with one of them is at $0\degree$. Once the beam angles are selected, FMO is performed using our optimization engine (Men *et al.*, 2009), which tries to optimize an energy function consisting only the $E_{Dose}$ term. The achieved final objective function value, denoted as $E_{Dose}^{FMO}$, is used to quantitatively evaluate the quality of the beam angle configurations, as

25   this $E_{Dose}$ is the only goal as to what is optimized in our BOO problem in addition to the goal of number of beams. The lower $E_{Dose}^{FMO}$ is, it is understood that the better the beam configuration is. In this example, it is found that the final objective value is $E_{Dose}^{FMO} = 1028$ for the angles generated by our BOO algorithm, a 14.2% decrease compared to $E_{Dose}^{FMO} = 1198$ for the equiangular case. The achieved lower dosimetric objective

30   function value indicates the better quality of the selected beam angles via our BOO algorithm.

One issue regarding the equiangular plan is that there are many ways to place the given number of beams depending on where the first beam is oriented. It has been found that the plan quality will considerably vary with the starting beam angle for the

35   equiangular beam orientations for prostate cancer (Potrebko *et al.*, 2007). On the other hand, varying the starting beam angle is also, to a certain extent, an optimization with respect to the beam orientations. To compare our BOO plan with all the equiangular plans, we have enumerated all the possible starting angles and hence all the equiangular plans, and compare our BOO plans with all of them. It is found that by adjusting the

40   starting beam angle, the $E_{Dose}^{FMO}$ values for this example case range from 1155 to 1211. Compare with the value of 1028 obtained from our optimized beam angles, the BOO algorithm generates beam configurations better than all possible equiangular beams indicated by the decrease of the $E_{Dose}^{FMO}$ values by 11.0~15.1%.

To better evaluate the treatment plan quality, a single number such as the final FMO objection function value $E_{Dose}^{FMO}$ is not sufficient. The plan quality has to be carefully examined in more clinical relevant ways. As such, for the same 7-beam prostate case, we first plot the dose color wash for the plan based on optimized angles and the corresponding one with equiangular beams starting at $0°$ in Fig. 3. The isodose lines are taken at $73.8 \times 1.1$, $73.8$, $73.8 \times 0.9$ and 30 Gy. We have also shown the DVHs in Fig. 4 corresponding to the two plans in Fig. 3. By examining the dose wash plots and the DVHs, it is found that the optimized beam angles attain a better PTV dose coverage and less hot spots than that with equiangular beams. For critical structures, the optimized beam angles lead to considerable improvement in terms of sparing dose at rectum and bladder, though paying a price of higher dose at femoral heads. This is due to the relatively low penalty coefficients $\beta$ for femoral heads compared with the other two critical structures.

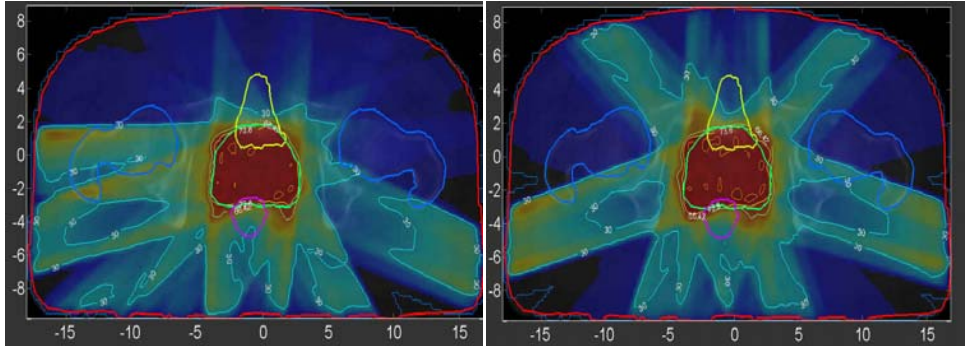

**Figure 3.** The dose color wash plots for the plan with 7 coplanar beams after BOO (left) and that for equiangular beams starting at $0°$ (right).

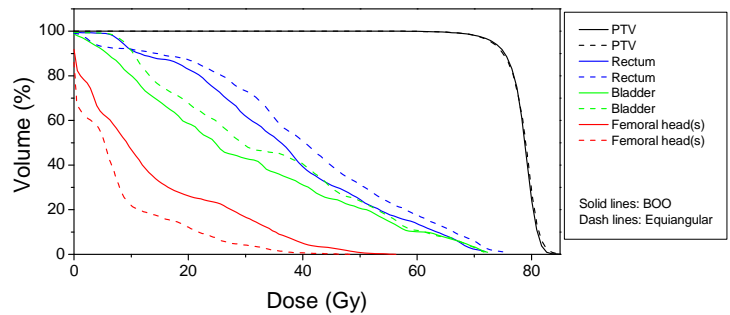

**Figure 4.** The DVH plot for the plan with 7 coplanar beams after BOO (solid lines) and that for equiangular beams starting at $0°$ (dash lines).

### 3.1.3 Different Number of Beams

For the same prostate case, we have also performed BOO with different target number of beams $N_A = 5~11$. For each given number of beams, our BOO algorithm gives the corresponding set of beam orientations. The selected beam angles are listed in Table 1. From this table, it is found that there exist some key beam orientations. By comparing the results under different $N_A$, apparently the beam angles $90°$, $125°$, $215°$, and $250°$ repeatedly present in the set of selected angles regardless of $N_A$, indicating the crucial role of these beam angles for a good treatment plan.

Once the beam angles are selected, FMO is then performed to obtain the final objective values $E_{Dose}^{FMO}$ based on the optimized beam orientations. For comparison purposes, FMO based on equiangular plans with all possible starting angles are also conducted. The results are summarized in Fig. 5, where the $E_{Dose}^{FMO}$ values are plotted as functions of the target number of beams for the optimized beam configuration and the equiangular configuration with a starting angle of $0°$. Meanwhile, a shaded area indicates the range of the $E_{Dose}^{FMO}$ values by varying the starting beam angle for the equiangular cases.

| $N_A$ | Selected beam angles |
|---|---|
| 5 | $90°, 125°, 145°, 215°, 250°$ |
| 6 | $90°, 105°, 125°, 175°, 215°, 250°$ |
| 7 | $90°, 105°, 125°, 145°, 175°, 215°, 250°$ |
| 8 | $90°, 105°, 125°, 145°, 175°, 215°, 250°, 270°$ |
| 9 | $70°, 90°, 105°, 125°, 145°, 175°, 215°, 250°, 270°$ |
| 10 | $70°, 90°, 105°, 125°, 145°, 175°, 215°, 240°, 250°, 270°$ |
| 11 | $70°, 90°, 105°, 125°, 145°, 175°, 215°, 240°, 250°, 270°, 325°$ |

**Table 1.** The beam angles selected by our BOO algorithm for a typical prostate case.

In general, the more angles used in IMRT, the FMO algorithm has more freedom during its optimization process, provided that the additional beams does not duplicate the role of any existing ones. This is the case for the optimized beam orientations. With our BOO algorithm, for each given number of beams, their orientations are effectively adjusted. Therefore, $E_{Dose}^{FMO}$ value monotonically decreases as a function of the target number of beams. On the other hand, for equiangular plans, parallel opposed beams exist in those cases with even number of beams. This is a well known situation of wasting beam angles in IMRT. In terms of objective values, $E_{Dose}^{FMO}$ at any even $N_A$ shows little decrease or even increase compared with the case at preceded odd $N_A$. Yet, if we only consider those odd number of beams, *i.e.* $N_A = 5, 7, 9, 11$, there is no waste of beam angles in equiangular plans. $E_{Dose}^{FMO}$ value hence decreases with $N_A$.

As for comparison between the optimized beam configurations and equiangular ones, the $E_{Dose}^{FMO}$ curve for the BOO cases is consistently lower than that for the equiangular ones, even lower than all the possible equiangular cases by placing starting angles differently. Moreover, in terms of the $E_{Dose}^{FMO}$ value, Fig. 5 indicates that a small number of beams with BOO can yield a plan better or comparable to equiangular plans with larger number of beams. Take the BOO case with $N_A = 6$ as an example, its $E_{Dose}^{FMO}$ value is less than that for $N_A = 9$ in the equiangular cases with a starting angle $0°$ and almost comparable to that for $N_A = 11$. Yet, the difference between $E_{Dose}^{FMO}$ values with and without BOO diminishes as more beams are used, indicating that the advantages of performing BOO becomes less profound.

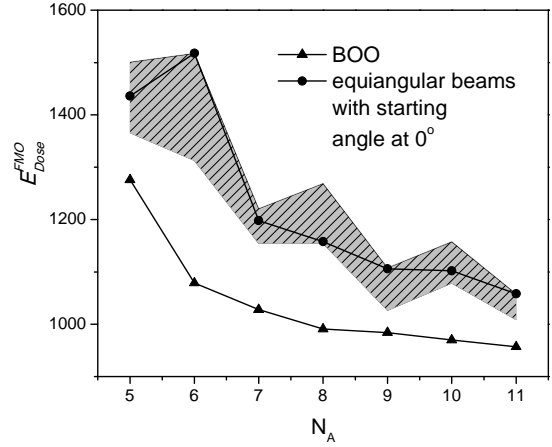

**Figure 5.** The FMO objective function values $E_{Dose}^{FMO}$ as a function of the target number of beams for optimized beam configurations and equiangular beam configurations. Shaded area indicates the range of the $E_{Dose}^{FMO}$ values by varying the starting beam angle for the equiangular cases.

5          For the proper number of beams used in IMRT, Fig .5 indicates that once the number of beams reaches a certain level, using more beams cannot improve the plan quality, provided that the beam configurations are optimized. In Fig. 5, though the $E_{Dose}^{FMO}$ decays as the number of beams increases in the BOO case, it tends to saturate when a large enough number of beams are used. For instance, from $N_A = 8$ to $N_A = 11$, $E_{Dose}^{FMO}$

10        decreases from 991 to 957, by only 3%, while 3 more beam angles are used. On the other hand, for the unoptimized beam configurations, such as in those equiangular cases, using more beams is certainly advantageous.

*3.2 All testing cases*

To systematically validate our BOO algorithms, we have performed studies on five clinical cases of prostate cancer (Case P1~P5). For each case, five to nine 6 MV coplanar beams are selected via our algorithms. In addition, one head-and-neck (HN) case (Case H1) with coplanar beams is also studied. Since treatment with coplanar beams is rarely used for HN cases, we only study BOO in one HN case to demonstrate the feasibility of

20        our BOO algorithm. This HN case contains two PTVs with prescription dose of 73.8 Gy to PTV1 and 54 Gy to a low dose target PTV2. PTV1 consists GTV expanded to account for both subclinical disease as well as daily setup errors and internal organ motion; PTV2 is a larger target that also contains high-risk nodal regions and is again expanded for

25        same reasons. Again, plans generated based on both optimized beam angles and equiangular beams with all possible starting angles are studied.

The resulted final energy function values after FMO for all cases are summarized in Table 2. Except in two cases (case P4 with $N_A = 7$ and case H1 with $N_A = 5$), the $E_{Dose}^{FMO}$ values for the optimized beam angles are lower than those for the equiangular configurations. For those cases where BOO yields a larger $E_{Dose}^{FMO}$ value, for instance the case P4 with $N_A = 7$, $E_{Dose}^{FMO}$ in BOO lies in the lower end of the range by varying the starting angle of the equiangular beam configurations. These results clearly indicate the effectiveness of our BOO algorithms.

| $N_A$ Case | 5 | 6 | 7 | 8 | 9 |
|---|---|---|---|---|---|
| P1 | **1773**/3591 [2636-4255] | **1725**/3222 [3136-4944] | **1606**/2391 [2123-2391] | **1581**/2700 [2239-2700] | **1407**/2102 [1828-2102] |
| P2 | **977**/1099 [1068-1137] | **908**/1066 [1066-1103] | **834**/897 [861-897] | **790**/862 [859-909] | **769**/795 [774-797] |
| P3 | **1276**/1436 [1366-1501] | **1079**/1518 [1313-1518] | **1028**/1198 [1155-1211] | **991**/1158 [1155-1269] | **984**/1106 [1064-1106] |
| P4 | **955**/1090 [972-1090] | **925**/1004 [1004-1068] | **837**/851 [824-913] | **810**/876 [856-971] | **774**/801 [777-801] |
| P5 | **1089**/1184 [1090-1332] | **996**/1206 [1125-1327] | **899**/973 [921-998] | **842**/1054 [922-1003] | **815**/843 [843-886] |
| H1 | **191**/185 [185-240] | **163**/238 [237-265] | **157**/181 [163-181] | **155**/189 [172-201] | **144**/152 [148-168] |

**Table 2.** The optimized energy function value $E_{Dose}^{FMO}$ based on different number of equiangular or optimized beam orientations in 5 coplanar prostate cases (P1~P5) and 1 coplanar HN case (H1). In each cell the number in bold font is for the optimized beam orientations, while the other one is for equiangular configuration starting at $0°$. Numbers in the parenthesis indicate the range of $E_{Dose}^{FMO}$ value by varying the starting angle of the equiangular configuration.

## 4. Discussion and Conclusions

In this paper, we have developed a new beam orientation optimization algorithm for IMRT treatment planning via adaptive $l_1$ minimization. By integrating the concept of sparsity into the optimization process and adaptively selecting weighting factors for each beam angle, we are able to identify the unimportant beam angles and gradually remove them without largely sacrificing the dosimetric objective. Such a process terminates when a given target number of beams is achieved. This algorithm has been systematically validated in the context of 5-9 coplanar beams for 5 prostate cases and 1 HN case. In one typical prostate case, the convergence properties of our algorithm, as well as how the beam angles are selected during our optimization process, are demonstrated. The resulted plan quality based on the optimized beam angles is also presented and compared with equiangular beam orientations. For all testing cases, FMO is performed based on the optimized beam configuration as well as on equiangular beams with all possible starting angles. The final FMO energy function value is used to compare the optimized beam orientations and the equiangular ones. It is found that, in all most all cases our BOO algorithm can lead to beam configurations which attain lower FMO energy function values than corresponding equiangular cases, indicating the effectiveness of our BOO algorithm.

Computation time is always a key component in evaluating the clinical feasibility of an algorithm. For our BOO algorithm, the computation is conducted on an Intel Xeon 2.27 GHz CPU with 8 GB memory using MATLAB for the purpose of proof of principle. Currently, the computational time is about 10 minutes for prostate cases and is about a few hours in head and neck cases due to their complicated anatomical geometry and many dosimetric objectives on different organs. It is expected that the optimization process will be considerably sped up when coded in more efficient computer language such as C, or when implemented on a GPU platform (Men *et al.*, 2009). In those circumstances, the computational time will not be a concern.

As a paper proposing the use of sparsity and adaptive $l_1$ minimization for solving BOO, we have only presented the results in the context of coplanar beams for simplicity. In real clinic, noncoplanar beams are also used in IMRT treatment, especially for HN patients. In principle, our BOO algorithm can also be applied for selecting beam orientations among a set of noncoplanar candidate beams. For this purpose, the dosimetric energy term $E_{Dose}$ will have to be modified to account for those noncoplanar beam candidates. If so, the dose deposition matrix $D$ will be considerably larger than those dealt in this paper due to the large number of candidate beams. It is expected that the computational time will be prolonged by the enlarged data size. Therefore, more effective implementation of our algorithm will be required.

The idea proposed in this paper is rather a general framework for beam orientation optimization as opposed to merely an algorithm. It actually provides us a lot of freedom to integrate more clinical constraints in the BOO process. In this paper, for simplicity, we only choose a quadratic one-sided voxel-based objective function $E_{Dose}$ to enforce the dosimetric objectives. Yet, the optimization process does not depend on the explicit form of $E_{Dose}$, as seen in Section 2.3. It is the sparsity idea and the adaptively selected weighting factors that leads to an optimized beam configurations. Therefore, one can easily incorporate more clinically relevant terms in $E_{Dose}$ for various clinical considerations. For instance, a total variation term (Zhu *et al.*, 2008) on the fluence map can be added in $E_{Dose}$ to enforce a desired degree of smoothness on the resulted fluence map to facilitate the leaf sequence optimization subsequent to the FMO. One can also substitute this $E_{Dose}$ by a term of another format, such as that in dose volume-based or equivalent uniform dose-based optimization (Das, 2009), to achieve BOO under dosimetric constraints with clinical or biological meanings.

Another important issue is the choice of the parameter $\mu$ in our algorithm. Generally speaking, the value of $\mu$ controls to what extent we would like to enforce the sparsity among beam angles. Apparently, we prefer a more sparse $\left\| x^\theta \right\|_2$ profile when we would like to choose less number of beam orientations. Therefore, a small $\mu$ value should be assigned when the target number of beams $N_A$ is small. In the testing cases studied in this paper, we have manually selected the $\mu$ value so that the best BOO results can be obtained. However, it is found that the selected $\mu$ values depend mainly on the number of beams $N_A$ for a given cancer site, but varying little from patient to patient. Thus, the $\mu$ values determined in our paper are practically usable in more clinical cases. This finding, however, is only based on our preliminary studies. The topic of how to determine the

5     parameter $\mu$ requires further study on a large patient group for each cancer site and will be address in our future work.

**Acknowledgements**

10    This work is supported in part by the University of California Lab Fees Research Program.

**Appendix**
**Derivation of Eq. (6)**

Observe that the energy $E_{Aux}[x]$ can be written as a summation of terms corresponding to different angles $\theta$, namely $E_{Aux}[x] = \sum_\theta E_{Aux}^\theta[x^\theta]$, where

$$E_{Aux}^\theta[x^\theta] = \frac{1}{2}\|x^\theta - g^\theta\|_2^2 + \lambda\mu_\theta\|x^\theta\|_2. \tag{A1}$$

It therefore suffices to consider the minimization for each $\theta$ term independently. To simplify notation, we denote $\hat{\mu} = \lambda\mu_\theta > 0$ and drop the superscript or subscript $\theta$ from hereon, wherever no confusion arises. Since this energy is convex, let us consider the optimality condition:

$$0 = (x - g) + \hat{\mu}\frac{x}{\|x\|_2}. \tag{A2}$$

It is straightforward to test that the solution to this equation is of the form $x = cg$, where $c$ is a constant. This leads to the equation

$$0 = c - 1 + \frac{\hat{\mu}}{\|g\|_2}\mathrm{sgn}(c), \tag{A3}$$

where $\mathrm{sgn}(.)$ is sign function. Eq. (A3) has only one solution at $c = 1 - \frac{\hat{\mu}}{\|g\|_2}$, under the condition $\hat{\mu} > \|g\|_2$. The corresponding solution that minimizes $E_{Aux}^\theta[x]$ is attained at $x_1^* = g(1 - \frac{\hat{\mu}}{\|g\|_2})$. Moreover, since $E_{Aux}^\theta[x]$ is not differentiable at $x_2^* = 0$, we have to consider this case separately.

Case 1) when $\hat{\mu} > \|g\|_2$, either $x_1^*$ or $x_2^*$ could be the minimizer of $E_{Aux}^\theta[x]$. However, it is not difficult to show that $E_{Aux}^\theta[x_1^*] \leq E_{Aux}^\theta[x_2^*]$. Hence $x_1^*$ minimizes the energy function. Case 2) when $\hat{\mu} < \|g\|_2$, only $x_2^*$ could be the minimizer of $E_{Aux}^\theta[x]$ and the solution is hence attained at $x_2^*$. In considering both cases, Eq. (6) follows.